\documentclass[10pt]{iopjournal}

\usepackage{amssymb,color,mciteplus,graphicx,subfigure}
\usepackage{calc,epsfig,epstopdf,mciteplus,bm,mathrsfs}
\usepackage{times}
\usepackage{float}
\usepackage{lipsum}
\usepackage{color}
\usepackage{amsmath}
\usepackage{ulem}
\usepackage{newtxtext,newtxmath}
\usepackage{url} 

\begin{document}

\articletype{Paper} 
	
\title{Dynamically Optimized Super-Robust Nonadiabatic Holonomic Quantum Gates Based on Superconducting Circuits}
	
\author{Hai Xu$^{1,2}$\orcid{0000-0003-0186-3898}, Wanchun Li$^3$, Tao Chen$^{4,5,*}$\orcid{https://orcid.org/0000-0001-5619-5781}, Kejin Wei$^{1,2}$ and Chengxian Zhang$^{1,2,*}$\orcid{0000-0003-4262-6055}}
	
\affil{$^1$School of Physical Science and Technology, Guangxi University, Nanning 530004, China}
	
	\affil{$^2$Guangxi Key Laboratory for Relativistic Astrophysics,
		School of Physical Science and Technology, Guangxi University, Nanning 530004, China}
	
	\affil{$^3$School of Science, Guilin University of Aerospace Technology, No.2, Jinji Road, Guilin, 541004, China}
	
		\affil{$^4$Key Laboratory of Atomic and Subatomic Structure and Quantum Control (Ministry of Education),
			School of Physics, South China Normal University, Guangzhou 510006, China}
	
		\affil{$^5$Guangdong Provincial Key Laboratory of Quantum Engineering and Quantum Materials, Guangdong-Hong Kong
			Joint Laboratory of Quantum Matter, Frontier Research Institute for Physics,
			South China Normal University, Guangzhou 510006, China}

	\affil{$^*$Author to whom any correspondence should be addressed.}

	\email{chentamail@163.com and cxzhang@gxu.edu.cn}


\begin{abstract}
Nonadiabatic holonomic quantum computation (NHQC) leverages non-Abelian geometric phases within a nonadiabatic framework to achieve fast and robust quantum gate operations. However, the practical implementation of NHQC is challenged by the imperfect control inherent in experimental environments. Ensuring deep suppression of control error is critical. In this work, we propose a dynamically optimized NHQC scheme to construct universal super-robust holonomic quantum gates. The proposed scheme is implemented by strategically designing a set of dynamically correcting pulses to achieve cyclic evolution, while ensuring that unwanted and disruptive dynamical phase elements, including previously neglected cross-coupling terms, are not accumulated. This constructed super-robust NHQC scheme efficiently safeguards the cyclic evolution process and makes the holonomic gate immune to control error by effectively correcting the error up to the fourth order. Furthermore, when integrated with decoherence-free subspace (DFS) encoding in superconducting quantum circuits, our scheme can achieve high-fidelity holonomic gates, and demonstrate robust resilience against both control errors and collective dephasing errors. Consequently, our work offers a promising pathway toward the realization of scalable and fault-tolerant holonomic quantum computation.
\end{abstract}

\section{Introduction}

Quantum computation has demonstrated capabilities for solving specific problems that are intractable in classical computation. Achieving robust control of quantum gates is critical for realizing fault-tolerant quantum computation \cite{Preskill2018,BhartiKishor_2022}. However, in practical physical implementations, quantum gates are inherently susceptible to various errors, resulting in significant infidelity. To address this challenge, a variety of error suppression techniques have been developed, including composite pulse sequences \cite{Yukihiro_2009,Torosov_2014,Gevorgyan_2021}, designing gates using geometric space curves \cite{DongWenzheng2021,Barnes2022,TangHoLun2023,NelsonHunter2023}, optimal control methods \cite{ZhuSLPRL2006,Carlini2012,DuYanXiong2016,Geng2016,DongYang2021}. Among these approaches, geometric quantum gates stand out as a stable and robust approach due to their reliance on the global properties of geometric phases. These phases are uniquely determined by the evolution path taken during the process and remain unaffected by specific details such as the speed of evolution \cite{berry1984quantal,Wilczek1984,Aharonov_1987,Anandan_1988,ZhuSLPRL2002,ZhuSLPRLUn2003}.

Geometric phases \cite{berry1984quantal,Wilczek1984}, which were initially studied in the adiabatic regime, were limited by the slow rate of cyclic evolution. To overcome this limitation, the concept was extended to include nonadiabatic cases \cite{Aharonov_1987,Anandan_1988,ZhuSLPRL2002,ZhuSLPRLUn2003}, thereby broadening its applicability to faster processes and a wider range of systems. As a result, proposals for nonadiabatic holonomic quantum computation (NHQC) \cite{zanardi1999,Sjoqvist_2012,XuGF_2012,XuGF2015,Herterich_2016,LiuBaoJie_2017,HongZhuoPing_2018,ZhaoPZ_2020,LiuBaoJie_2020,JiLiNa_2022,LiangYan_2022,LiuBaoJie_2023,ZhaoPZ_2023,YuXiaoDong_2023,zhang2023geo,liang2023non}, based on non-Abelian nonadiabatic geometric phases, have emerged, further expanding the potential of geometric phases in quantum computing. So far, the holonomic gates have been experimentally verified in various quantum platforms, such as superconducting quantum circuits \cite{abdumalikov2013,XuYun_2018,YanTongxing_2019,LiSai_2021exper}, trapped ions \cite{AiMingZhong_2022}, nuclear
magnetic resonance \cite{FengGuanru_2013,LiHang_2017,ZhuZhennan_2019}, and nitrogen-vacancy centers in diamond \cite{ArroyoCamejo_2014,ZuChong_2014,ZhouBrian_2017,Sekiguchi_2017,IshidaNaoki_2018,DongYang_2021}. It is recognized that perfect control cannot be achieved in the experimental environment, and the holonomic gates remain susceptible to control errors \cite{ZhengShiBiao_2016,JingJun_2017}.

To address the error correction challenge in NHQC, various theoretical approaches have been proposed, such as composite pulses \cite{ZhuZhennan_2019,XuGuoFu_2017}, dynamical decoupling \cite{XuGuoFu_2018,SekiguchiYuhei_2019,ZhaoPengZhi_2021}, pulse optimization \cite{YanTongxing_2019,LiuBaoJie_2019plug,ZhangFeihao_2019,LiSai_2020fast,Ai_MingZhongPRApp20,ChenTao_2020}, and dynamically corrected gates \cite{LiSai_2021,LiuBaoJie_2021super,HeZhiCheng_2021}. However, the enhancement of gate robustness often comes at the expense of the need for specific pulse shapes \cite{LiSai_2020fast,Ai_MingZhongPRApp20}. Moreover, it has been demonstrated that conventional NHQC schemes \cite{Sjoqvist_2012, Herterich_2016,HongZhuoPing_2018} are susceptible to extra dynamical phases induced by cross-coupling with states outside the computational subspace, which ultimately limits gate robustness. Although previous dynamically corrected NHQC schemes \cite{LiSai_2021exper,LiSai_2021,LiuBaoJie_2021super} have introduced super-robust conditions that effectively eliminate cross-coupling impacts and enhance holonomic gate robustness, there remains potential for improvement due to the passive imposition of additional geometric constraints. Consequently, there is an evident necessity for the further development of super-robust NHQC protocols that avoid additional burdens and prevent the accumulation of unnecessary disruptive dynamical phases.

In this paper, we propose a dynamically optimized NHQC scheme (OP-NHQC) to construct universal super-robust holonomic quantum gates. The proposed scheme is implemented by strategically designing a set of dynamically correcting pulses to achieve cyclic evolution, while ensuring that unwanted and disruptive dynamical phase elements, including previously neglected cross-coupling terms, are not accumulated. More importantly, the constructed super-robust NHQC scheme eliminates the need for additional complex geometric controls, relying instead on simple resonant interactions. Consequently, this approach efficiently safeguards the cyclic evolution process and makes the holonomic gate immune to control error by effectively correcting it up to the fourth order. Numerical simulation results demonstrate that our OP-NHQC scheme significantly enhance the gate robustness compared to the conventional single-loop NHQC schemes \cite{Sjoqvist_2012, Herterich_2016,HongZhuoPing_2018}, and it also outperforms both the dynamically corrected NHQC (DC-NHQC) schemes \cite{LiSai_2021exper,LiSai_2021,LiuBaoJie_2021super} and the composite NHQC (C-NHQC) schemes \cite{ZhuZhennan_2019,XuGuoFu_2017}. Finally, we further demonstrate that our scheme is highly compatible with DFS encoding \cite{ZanardiP_1997,DuanLu-Ming_1997,LidarDA_1998} in superconducting quantum circuits. This compatibility enables our approach to simultaneously mitigate both control error and collective dephasing error, thereby facilitating the implementation of high-fidelity holonomic gates.

\begin{figure*}[t]
	\begin{center}
		\includegraphics[width=0.6\columnwidth]{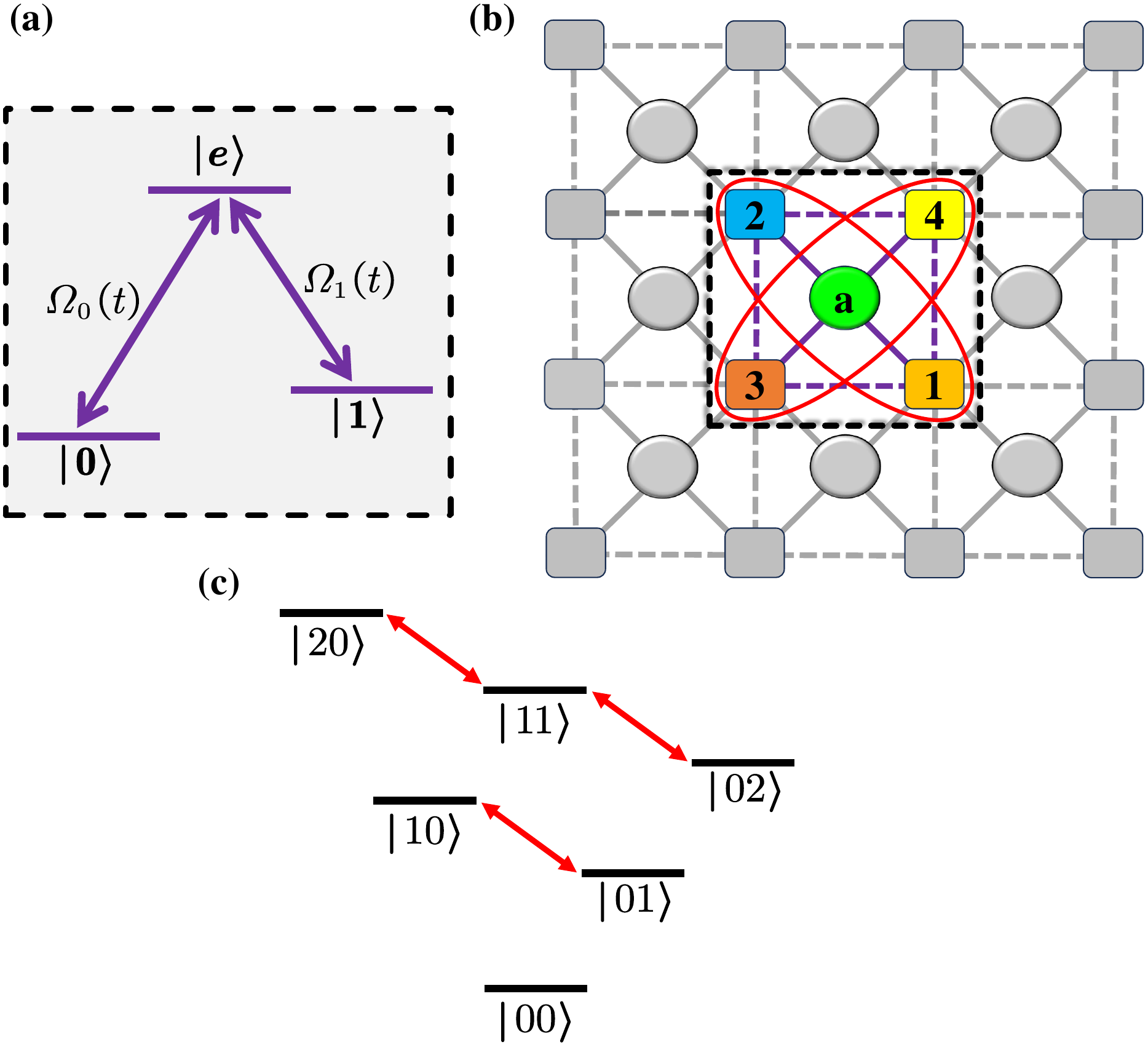}
		\caption{(a) The $\Lambda$-type three-level system, with resonant driving between the $|0\rangle\leftrightarrow |e\rangle$ and $|1\rangle\leftrightarrow |e\rangle$ transitions. (b) The coupling structure for realizing the DFS encoding, where each square denotes a superconducting transmon qubit. Here, the physical qubits with different frequencies are represented by the (orange, gold, yellow and blue) colored squares. While the circle with green colors is regarded as an auxiliary physical qubit. (c) The energy structure for two parametrically tunable coupled transmons, where we use the single- and two-excitation subspaces to realize the single- and two-logical-qubit holonomic gates. respectively.}\label{fig1}
	\end{center}
\end{figure*}

\section{Conventional holonomic gates}
To begin with, we first introduce the realization of the conventional single-loop NHQC scheme \cite{Sjoqvist_2012, Herterich_2016,HongZhuoPing_2018}, which is based on a general $\Lambda$-type three-level quantum system. As shown in Fig.~\ref{fig1}(a), the system is resonantly driven by two different external pulses with the amplitudes $\Omega_0(t)$ and $\Omega_1(t)$. $\phi_0(t)$ and $\phi_1(t)$ are the corresponding phases, respectively. Hereafter, we set $\hbar = 1$. Then, the Hamiltonian of the three-level system in the interaction picture is given by
\begin{equation}
	\begin{aligned}
		\mathcal{H}(t) =& \left[\Omega_0(t)e^{-i\phi_0(t)}|0\rangle + \Omega_1(t)e^{-i\phi_1(t)}|1\rangle \right]\langle e| + \textrm{H.c.}\\
		=& \Omega(t) e^{-i\phi_0(t)}|b\rangle\langle e|+ \textrm{H.c.}.
	\end{aligned}
	\label{Eq:HI}
\end{equation}
Here, $\{|0\rangle, |1\rangle\}$ are the computational basis states, $|e\rangle$ denotes the auxiliary state, and $|b\rangle = \sin(\theta/2)|0\rangle - \cos(\theta/2)e^{i\phi}|1\rangle$ is the bright state with $\phi = \phi_0-\phi_1+\pi$.  Except for the bright state, $\mathcal{H}(t)$ also has a dark state $|d\rangle=-\cos(\theta/2)e^{-i\phi}|0\rangle-\sin(\theta/2)|1\rangle$, which is decoupled from the quantum dynamics. The amplitudes above have been parameterized as $\Omega(t) = \sqrt{\Omega^2_0(t)+\Omega^2_1(t)}$, $\tan(\theta/2)=\Omega_0(t)/\Omega_1(t)$. Note that $\theta$ is fixed for a specific rotation (see Eq.~\eqref{Eq:UU}). An arbitrary evolution of the system can be formulated as $U\left(t,0\right)=\mathcal{T}\exp\left(-i\int_{0}^{t}H\left(t^{\prime}\right)dt^{\prime}\right)=\sum_{k=d, b, e}|\psi_{k}\left(t\right)\rangle\langle\psi_{k}\left(0\right)|$, where $\mathcal{T}$ is the time-ordering operator. And the evolution state is defined as $|\psi_{k}\left(t\right)\rangle=U\left(t,0\right)|k\rangle$. It is therefore possible to utilize the Hamiltonian parameters $\Omega\left(t\right)$ and $\phi_{0}\left(t\right)$ to engineer a holonomic evolution. Furthermore, to construct nonadiabatic holonomic quantum gates, the conventional NHQC schemes \cite{Sjoqvist_2012, Herterich_2016,HongZhuoPing_2018} require satisfying the conditions for cyclic evolution and parallel transport, that is,
\begin{eqnarray} \label{EqPT}
	&&\!\sum_{k=d,b}|\psi_{k}\left(\tau\right)\rangle\langle\psi_{k}\left(\tau\right)|=\sum_{k=d,b}|\psi_{k}\left(0\right)\rangle\langle\psi_{k}\left(0\right)|,  \notag \\
	&&\langle\psi_{k}\left(t\right)|H\left(t\right)|\psi_{l}\left(t\right)\rangle=d_{kl}\left(t\right)=0, \ \ k, l=d,b.
\end{eqnarray}
Due to the dark state $|d\rangle$ being decoupled from the system evolution at all times, i.e., $H\left(t\right)|d\rangle=0$, $d_{dd}\left(t\right)$, $d_{db}\left(t\right)$, and $d_{bd}\left(t\right)$ are all equal to zero. In this way, the above parallel-transport condition is reduced to
\begin{equation}
	d_{bb}\left(t\right)=\langle\psi_{b}\left(t\right)|H\left(t\right)|\psi_{b}\left(t\right)\rangle = 0.
\end{equation}
Obviously, it only corresponds to the partial elimination of dynamical phase elements during the evolution. In the subsequent discussion, we show that the non-zero cross-coupling term $d_{be}\left(t\right)=\langle\psi_{b}\left(t\right)|H\left(t\right)|\psi_{e}\left(t\right)\rangle$, which is also associated with dynamical phase elements, is neglected in the conventional NHQC schemes  \cite{Sjoqvist_2012, Herterich_2016,HongZhuoPing_2018}, and removing it is crucial for effectively improving gate robustness.

\begin{figure*}[tbp]
	\begin{center}
		\includegraphics[width=0.6\columnwidth]{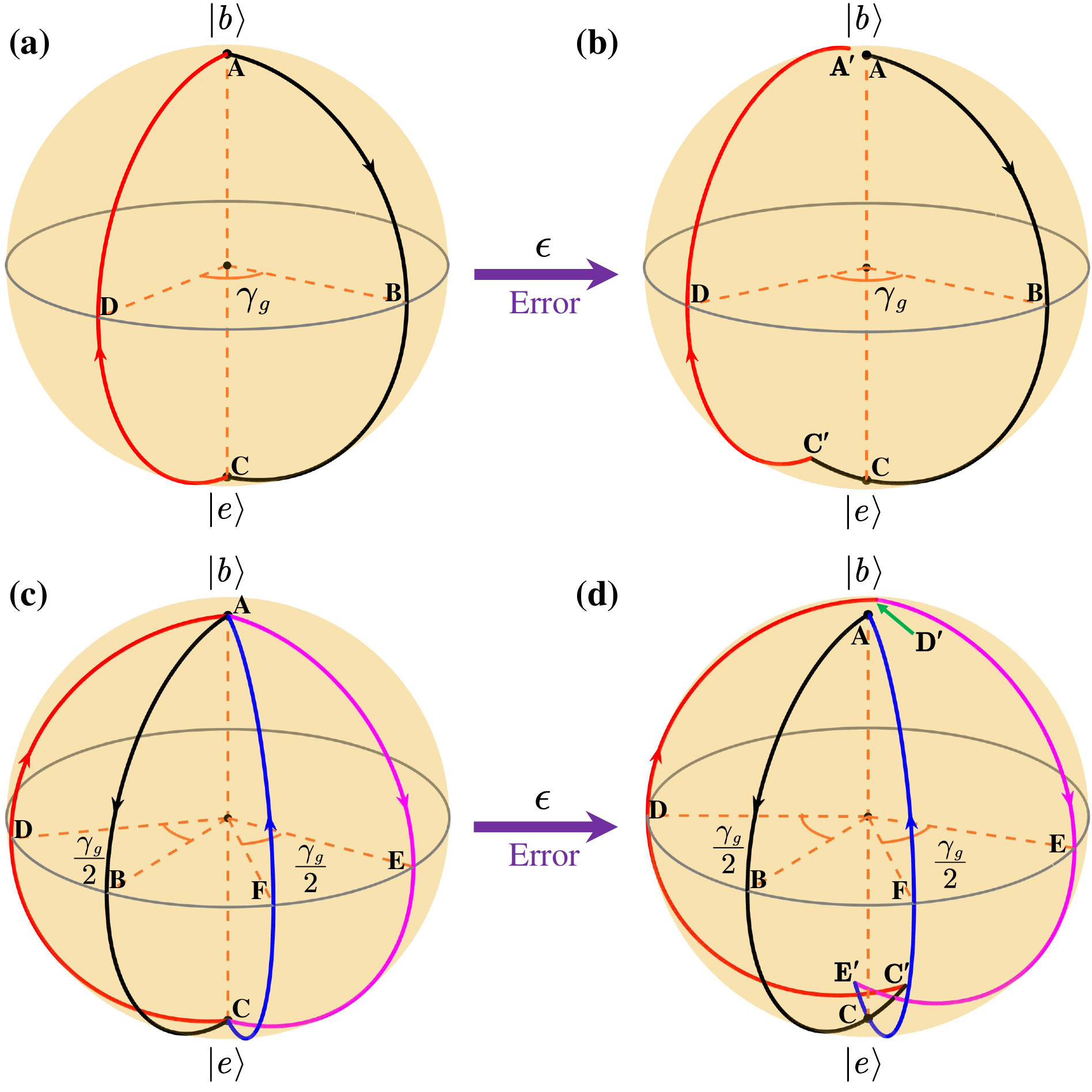}
		\caption{The evolution trajectories for holonomic quantum gates in Eq.~\eqref{Eq:UU} and Eq.~\eqref{Eq:UU_OPNHQC} with $\gamma_g = \pi/2$. Without considering the control error $\epsilon$, the trajectories in (a) and (c) denote the evolution path  for the conventional NHQC and OP-NHQC schemes, respectively. In (a), the trajectory is along $A\rightarrow B\rightarrow C\rightarrow D\rightarrow A$, while  in (c), it is along $A\rightarrow B\rightarrow C\rightarrow D\rightarrow A\rightarrow E\rightarrow C\rightarrow F\rightarrow A$. (b) and (d) are the corresponding cases for (a) and (c), respectively, considering  the presence of control error.} \label{fig2}
	\end{center}
\end{figure*}

The conventional single-loop NHQC schemes can be implemented by dividing the evolution into two segments
\begin{eqnarray}  \textbf{\label{Eq:Con_NHQC_Sche}}
	&&\!\!\!\!t\in\left[0, \tau_{1}\right], \ \ \int_{0}^{\tau_{1}}\Omega(t)dt= \frac{\pi}{2}, \ \ \phi_{0}(t)=\phi_{0}^{\prime}, \\
	&&\!\!\!\!t\in\left[\tau_{1},\tau\right], \ \ \int_{\tau_{1}}^{\tau}\Omega(t)dt= \frac{\pi}{2}, \ \ \phi_{0}(t)=\phi_{0}^{\prime}+\pi-\gamma_{g}.	
\end{eqnarray} 
In these two segments, the corresponding evolution operators are $ U_{1}= \mathrm{exp}(-i\int_{0}^{\tau_{1}}\mathcal{H}(t)dt)$ and $ U_{2}= \mathrm{exp}(-i\int_{\tau_{1}}^{\tau}\mathcal{H}(t)dt)$.  As shown in Fig.~\ref{fig2}(a), during the first half of the evolution, the evolution starts from the north pole and travels to the south pole along a given longitude. At the end of the second half of the evolution, it goes back to the north pole along another longitude ($\gamma_{g}$ away from the original one) to finish the cyclic evolution. Finally, $|b\rangle$ acquires a pure geometric phase $\gamma_g$. Meanwhile, during the whole process,  the dark state $|d\rangle$ does not change, because it always decouples from the system. Then, in the dressed states basis spanned by $\{|b\rangle, |d\rangle\}$, the holonomic gate is thus
\begin{eqnarray}\label{UU1}
	U = U_{2} U_{1}=|d\rangle \langle d| + e^{i\gamma_g}|b\rangle \langle b|.
\end{eqnarray}
On the other hand, one can easily convert the holonomic gate into the computational basis spanned by $\{|0\rangle, |1\rangle\}$, i.e.,
\begin{equation}
	\begin{aligned}
		U\left(\theta,\phi,\gamma_{g}\right) &=e^{i{\frac{ \gamma_g}{2}}}\left(\begin{array}{cc}
			\cos \frac{\gamma_{g}}{2}-i \sin \frac{\gamma_{g}}{2} \cos \theta & -i \sin \frac{\gamma_{g}}{2} \sin \theta e^{-i \phi} \\
			-i \sin \frac{\gamma_{g}}{2} \sin \theta e^{i \phi} & \cos \frac{\gamma_{g}}{2}+i \sin \frac{\gamma_{g}}{2} \cos \theta
		\end{array}\right) \\
		& =e^{i{\frac{ \gamma_g}{2}}}e^{-i{\frac{ \gamma_g}{2}}\mathbf{n}\cdot \boldmath{\sigma}},
	\end{aligned}
	\label{Eq:UU}
\end{equation}
where $\textbf{n}=(\sin\theta\cos\phi,\sin\theta\sin\phi,\cos\theta)$, and $\mathbf{\sigma}=(\sigma_x, \sigma_y, \sigma_z)$  with $\sigma_{x, y, z}$ representing the Pauli matrices in the computational basis. Then, by appropriately modulating the axis $\textbf{n}$ and the phase $\gamma_{g}$,  one can realize arbitrary single-qubit holonomic quantum gates.

In experiments, the $\Lambda$-type system would easily suffer from the control amplitude errors, which  can be induced due to the imprecise control of the microwave field. Thus, the amplitudes turn to be $\Omega_{0}(t)\rightarrow (1+\epsilon_{0}(t))\Omega_{0}(t)$ and $\Omega_{1}(t)\rightarrow (1+\epsilon_{1}(t))\Omega_{1}(t)$. Here, $\epsilon_{0}(t)$ and $\epsilon_{1}(t)$ denote the control amplitude error for $\Omega_{0}(t)$ and  $\Omega_{1}(t)$,  and they can fluctuate along with time. 
In this work, we assume $|\epsilon_{0}(t) |$, $|\epsilon_{1}(t) |\ll 1$, namely, they are regarded as random perturbations.  On the other hand, when these two errors are small enough, their  effects on the Hamiltonian can be approximated as $\mathcal{H}^{\epsilon}(t)=(1+\epsilon(t))\mathcal{H}(t)$, i.e., the driving amplitude $\Omega(t)$ with an additional error fraction of $\epsilon(t)\Omega(t)$ (for details, see Appendix \ref{AP3}). Typically, the error $\epsilon(t)\Omega(t)$  would lead to over-rotation issue, due to the fact that the pulse area related to the rotation angle turns to be $\int_{0}^{\tau}(1+\epsilon(t))\Omega(t)dt$ \cite{Bando.13}. Absolutetly, this hinders the realization of large-scale holonomic quantum computation.

In practice,  it is convenient to consider $\epsilon(t) $ as a constant value (i.e., $\epsilon(t)\equiv \epsilon$) in the Hamiltonian instead of  a time dependent random fluctuation. In this way, it is convenient to study the evolution operator and the error effect analytically using the perturbation theory. This assumption is reasonable considering the state-of-the-art experimental techniques, where the gate operation time is much shorter than the typical time scale of the stochastic error that varies along with time. In fact, such error model has been used in the recent related experiments in superconducting circuits \cite{XuYun_2018,LiSai_2021exper}. Nevertheless, the time dependent property of the errors is discussed in detail in Appendix \ref{AP3}.

The evolution operator with error can be expressed as  $U^{\epsilon}_{\rm{con}}(t)=e^{-i\int_{0}^{t}\mathcal{H}^\epsilon(t')dt'}$. In the dressed state basis $\{|b\rangle, |d\rangle\}$, it has the form as
\begin{equation}
	\begin{aligned}
		U^{\epsilon}_{\rm{con}} = |d\rangle \langle d| +\left( \cos^2\frac{\mu\pi}{2}+  \sin^2\frac{\mu\pi}{2}e^{i\gamma_g}  \right)|b\rangle \langle b|,
	\end{aligned}
	\label{Eq:UeNHQC}
\end{equation}
where $\mu = 1+\epsilon$. Based on this error operator, we can further expand the gate fidelity \cite{WangXiaoguang_2009} as 
\begin{equation}
	\begin{aligned}
		F_{\rm{con}}&=\frac{|\rm{Tr}(U^\dagger U^\epsilon)|}{|\rm{Tr}(U^\dagger U)|}\\&=\frac{1}{2}\left|1+\cos^2\frac{\mu\pi}{2}e^{-i\gamma_g}+\sin^2\frac{\mu\pi}{2}\right|\\
		&\approx 1-\epsilon^2\pi^2(1-\cos\gamma_g)/8.
	\end{aligned}
	\label{Eq:FUeNHQC}
\end{equation}
The conventional holonomic gate is immune to leading-order error, implying that it can suppress the control error $\epsilon$ only up to the second order. Moreover, although composite pulse control has been further incorporated into the construction of conventional holonomic gates, the resulting composite NQHC schemes \cite{ZhuZhennan_2019,XuGuoFu_2017} can only suppress the control error $\epsilon$ to the same level (for details, see Appendix \ref{AP1}). It is evident that the conventional NHQC gates and composite NHQC gates fall short in terms of enhanced resilience. Addressing this issue is key for the practical application of these gates.

\begin{figure*}[tbp]
	\begin{center}
		\includegraphics[width=0.7\columnwidth]{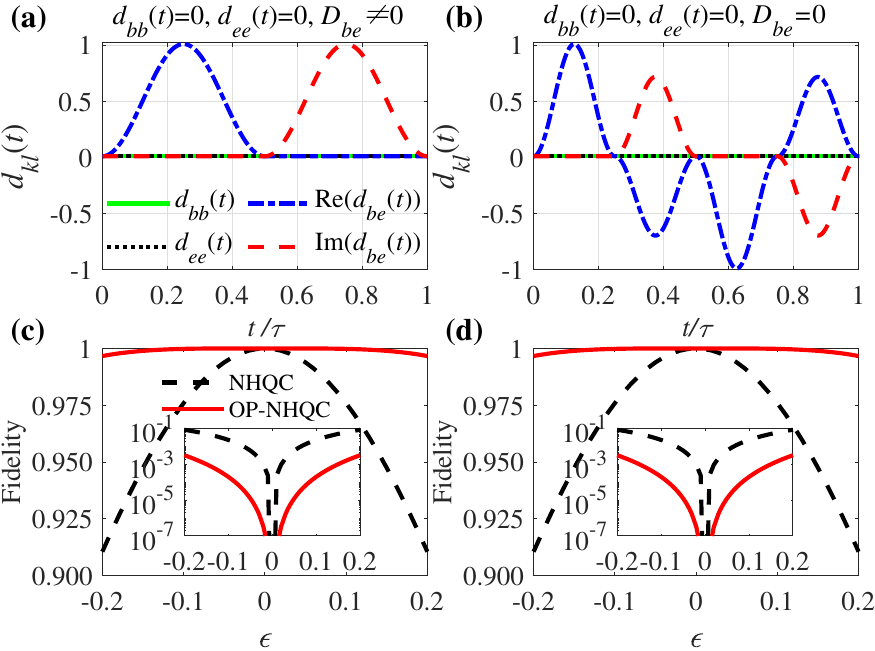}
		\caption{The dynamical phase accumulated over time evolution for (a) the conventional NHQC scheme and (b) our OP-NHQC scheme, with the holonomic $X/2$ gate taken as an example. The robustness comparison between the conventional NHQC scheme and our OP-NHQC scheme for holonomic $X/2$ gate in (c) and $S$ in (d).}\label{fig2DPhase}
	\end{center}
\end{figure*}

Next, we will focus on investigating the impact of dynamical phases accumulation during cyclic evolution on the error resilience of holonomic quantum gates. As illustrated in Fig.~\ref{fig2DPhase}(a), the dynamical phase element $d_{bb}\left(t\right)$ (see green line) is observed to be zero at all times for both the conventional NHQC scheme. Here, the pulse shape of the driving field is set as $\Omega\left(t\right)=\Omega_{m}\sin^{2}\left(\frac{\pi t}{\tau}\right)$ with $\Omega_{m}=1$ for simplicity to clearly visualize the changing process of dynamical phase elements during cyclic evolution. But, by calculating the accumulated phase in the full space, it can be found that the phase used for gate design in it is not purely geometric, due to the presence of non-zero cross-coupling term $d_{be}\left(t\right)$. As a result, as shown in Fig.~\ref{fig2}(b), the intrusion of control error causes the single-loop geometric evolution used to construct the conventional holonomic quantum gate to fail in closing cyclically. Consequently, this nonzero cross-coupling inevitably introduces an additional dynamical phase, thereby directly compromising the robustness of holonomic gates, as shown in Figs.~\ref{fig2DPhase}(c) and~\ref{fig2DPhase}(d).

\section{Dynamically optimized holonomic gates\label{Sec:DOPNHQG}}

To fully cancel out dynamical phase elements, the above-mentioned condition of parallel transport in Eq.~\eqref{EqPT} needs to be extended to include
\begin{equation} \label{EqCon1}
	D_{be} = \int_0^\tau d_{be}(t) d t=0.
\end{equation}
In this way, the extended condition enables the elimination of the dynamical element arising from nonzero cross-coupling term in the design of strong error-resilient holonomic gates, ensuring that the accumulated dynamical phases sum to zero after cyclic evolution.
Here, building on the extended parallel-transport condition described above, we propose a dynamically optimized NHQC scheme. The core idea behind the dynamically corrected gate adopted here is to replace a simple pulse with a sequence of elementary pulses \cite{Kestner.14,WangX.14}. These two pulse sets are, in fact, equivalent. Moreover, the noise in question is self-compensated and eliminated throughout the evolution process. 

The strategy employed here is the construction of a robust composite pulse sequence, in which the resulting new evolution operator $U'_2$ is shown to be equivalent to $U_2$. To achieve this, a carefully designed three-pulse sequence is proposed to replace the single pulse previously used to implement $U_2$. Conversely, the first-segment evolution $U_1$, as defined in Eq.~\eqref{Eq:Con_NHQC_Sche}, remains unaltered. 
For the purpose outlined above, the Hamiltonian control parameters are to be designed according to the following form
\begin{eqnarray}
	&&\!\!\!\!\!\!\!\!\!\!\!\!\!\!\!\!t\in\left[0,\tau_{1}\right],  \int_{0}^{\tau_{1}}\Omega(t)dt= \frac{\pi}{2}, \ \ \phi_{0}(t)=\phi_{0}^{\prime},  \label{Opulse1} \\
	&&\!\!\!\!\!\!\!\!\!\!\!\!\!\!\!\!t\in\left[\tau_{1},\tau_{2}\right],  \int_{\tau_{1}}^{\tau_{2}}\Omega(t)dt= \frac{\pi}{2}, \ \ \phi_{0}(t)=\phi_{0}^{\prime}+\pi-\frac{\gamma_{g}}{2}, \label{Opulse2}\\
	&&\!\!\!\!\!\!\!\!\!\!\!\!\!\!\!\!t\in\left[\tau_{2},\tau_{3}\right],  \int_{\tau_{2}}^{\tau_{3}}\Omega(t)dt= \frac{\pi}{2}, \ \ \phi_{0}(t)=\phi_{0}^{\prime}+\pi-\gamma_{g}, \label{Opulse3}\\
	&&\!\!\!\!\!\!\!\!\!\!\!\!\!\!\!\!t\in\left[\tau_{3},\tau\right],  \int_{\tau_{3}}^{\tau}\Omega(t)dt= \frac{\pi}{2}, \ \ \phi_{0}(t)=\phi_{0}^{\prime}+2\pi-\frac{3\gamma_{g}}{2}.
	\label{Opulse4}
\end{eqnarray}
Based on the above parameter settings, our dynamically optimized holonomic gate in the computational basis of $\{|0\rangle, |1\rangle\}$ can be solved as
\begin{equation}
	\begin{aligned}
		U_{\rm{op}} & = U_{2}^{\prime} U_{1}=e^{i{\frac{ \gamma_g}{2}}}e^{-i{\frac{ \gamma_g}{2}}\mathbf{n}\cdot \mathbf{\sigma}},
	\end{aligned}
	\label{Eq:UU_OPNHQC}
\end{equation}
which has the same form in Eq.~\eqref{Eq:UU}. Therefore, we can use this scheme to realize universal holonomic gates. As illustrated in Fig.~\ref{fig2}(c), we plot the evolution process for our dynamically optimized NHQC scheme without considering the effect of error. The black and red lines correspond to the two pulses in Eqs.~\eqref{Opulse1} and~\eqref{Opulse2}, which can drive the state $|b\rangle$ to accumulate a $\gamma_{g}/2$ geometric phase in the interval from $0$ to $\tau_{2}$. Similarly, the purple and blue lines, corresponding to the two pulses in Eqs.~\eqref{Opulse3} and~\eqref{Opulse4}, continue to drive the state $|b\rangle$ to accumulate a geometric phase of $\gamma_{g}/2$ during $\tau_{2}$ to $\tau$. This configuration guarantees that the geometric phase of $\gamma_{g}$ can be obtained after a full cycle evolution, which in turn enables the errors to be self-canceling.

It should be noted that although a set of dynamically correcting pulses is introduced, no additional dynamical phase elements accumulate throughout the evolution process because the evolution trajectory always remains aligned with the longitude. As illustrated in Fig.~\ref{fig2DPhase}(b), it can be observed that the dynamical phase element $d_{bb}\left(t\right)$ is consistently zero at all times. Unlike the conventional NHQC and composite NHQC schemes, our dynamically optimized NHQC scheme additionally satisfies the extended condition in Eq.~\eqref{EqCon1}, i.e., $D_{be}=0$. This ensures the elimination of the destabilizing factor, namely the cross-coupling term, that disrupts the robustness of the holonomic gate. As a result, even in the presence of control error $\epsilon$, the evolution used to achieve dynamically optimized holonomic gates remains cyclic and approximately closed, as shown in Fig.~\ref{fig2}(d).

\begin{table*}[tbp]
	\centering
	\tabcolsep=0.85cm
	\renewcommand\arraystretch{2.0}
	\caption{The robustness comparison with the gate infidelity of conventional NHQC, composite NHQC (C-NHQC), dynamically corrected NHQC (DC-NHQC) and our dynamically optimized NHQC (OP-NHQC).}
	\begin{tabular}{cccc}
		\hline\hline
		Types & Gate time & Infidelity  & References \\
		\hline
		NHQC  & $\pi/\Omega_{m}$ & $\epsilon^2\pi^2(1-\cos\gamma_g)/8$ &\cite{Sjoqvist_2012, Herterich_2016,HongZhuoPing_2018} \\
		C-NHQC  & $2\pi/\Omega_{m}$  &$\epsilon^{2}\pi^{2}\sin^{2}\frac{\gamma_{g}}{4}\cos^{2}\frac{\gamma_{g}}{2}$ & \cite{ZhuZhennan_2019,XuGuoFu_2017} \\
		DC-NHQC  & $2\pi/\Omega_{m}$  &$\epsilon^4\pi^4(1-\cos\gamma_g)/32$ & \cite{LiSai_2021exper,LiSai_2021,LiuBaoJie_2021super} \\
		OP-NHQC & $2\pi/\Omega_{m}$ & $\epsilon^4\pi^4(1-\cos\gamma_g)/64$ & This work \\
		\hline\hline
	\end{tabular}
	\label{table1}
\end{table*}

Furthermore, theoretical calculations show that the evolution operator for our dynamically optimized NHQC scheme in the dressed states basis $\{|b\rangle, |d\rangle\}$ changes from the ideal case to
\begin{equation}
	\begin{aligned}
		&U^{\epsilon}_{\rm{op}} = |d\rangle \langle d|
		+e^{i\gamma_{g}}\left [1-2i\sin\frac{\gamma_{g}}{2}\left(\cos^{4}\frac{\mu\pi}{2}e^{-i\frac{\gamma_{g}}{2}}+\frac{1}{4}\sin^{2}\mu\pi \right)  \right ]|b\rangle \langle b|.
	\end{aligned} \label{Eq:UeOPNHQC}
\end{equation}
The gate fidelity can then be expanded as
\begin{equation}
	\begin{aligned}
		F_{\rm{op}} =&\frac{1}{2}\left|2-2i\sin\frac{\gamma_{g}}{2}\left(\cos^{4}\frac{\mu\pi}{2}e^{-i\frac{\gamma_{g}}{2}}+\frac{1}{4}\sin^{2}\mu\pi\right)  \right| \\
		\approx & 1-\epsilon^4\pi^4(1-\cos\gamma_g)/64.
	\end{aligned}
	\label{Eq:FUeOPNHQC}
\end{equation}
The result shows that, compared to the conventional NHQC, our dynamically optimized NHQC can effectively enhances the error resilience of the holonomic gate, and suppress the control error $\epsilon$ up to the fourth order. The numerical simulations shown in Figs.~\ref{fig2DPhase}(c) and~\ref{fig2DPhase}(d) also validate this, with the holonomic $X/2$ and $S$ gates serving as representative examples.

On the other hand, the previous dynamically corrected NHQC schemes \cite{LiSai_2021exper,LiSai_2021,LiuBaoJie_2021super} also incorporate similar extended condition in Eq.~\eqref{EqCon1}, in the construction of holonomic gates. However, unlike both our dynamically optimized NHQC scheme and the conventional NHQC scheme, which maintain $d_{bb}\left(t\right)=0$ at all times, the dynamically corrected NHQC schemes only ensure that the integral of $d_{bb}\left(t\right)$ vanishes at the final time of the cyclic evolution, i.e., $\int_0^\tau d_{bb}(t) d t=0$. The details is shown in Appendix \ref{AP2}. This different results in the dynamically corrected NHQC schemes sacrifice some gate robustness. Therefore, although the dynamically corrected NHQC schemes show an improvement in gate robustness compared to the conventional NHQC schemes, it still falls short when compared to our current scheme. As shown in Table.~\ref{table1},
we compare the gate infidelity between the conventional NHQC, composite NHQC, dynamically corrected NHQC and our dynamically optimized NHQC. It is found that our dynamically optimized NHQC scheme performs the best, and the infidelity is only half of the dynamically corrected NHQC scheme. This finding is further substantiated by the numerical simulation depicted in Figs.~\ref{FigDPhase}(e) and~\ref{FigDPhase}(f) (see Appendix \ref{AP2}).

\begin{figure}[tbp]
	\begin{center}
		\includegraphics[width=0.65\columnwidth]{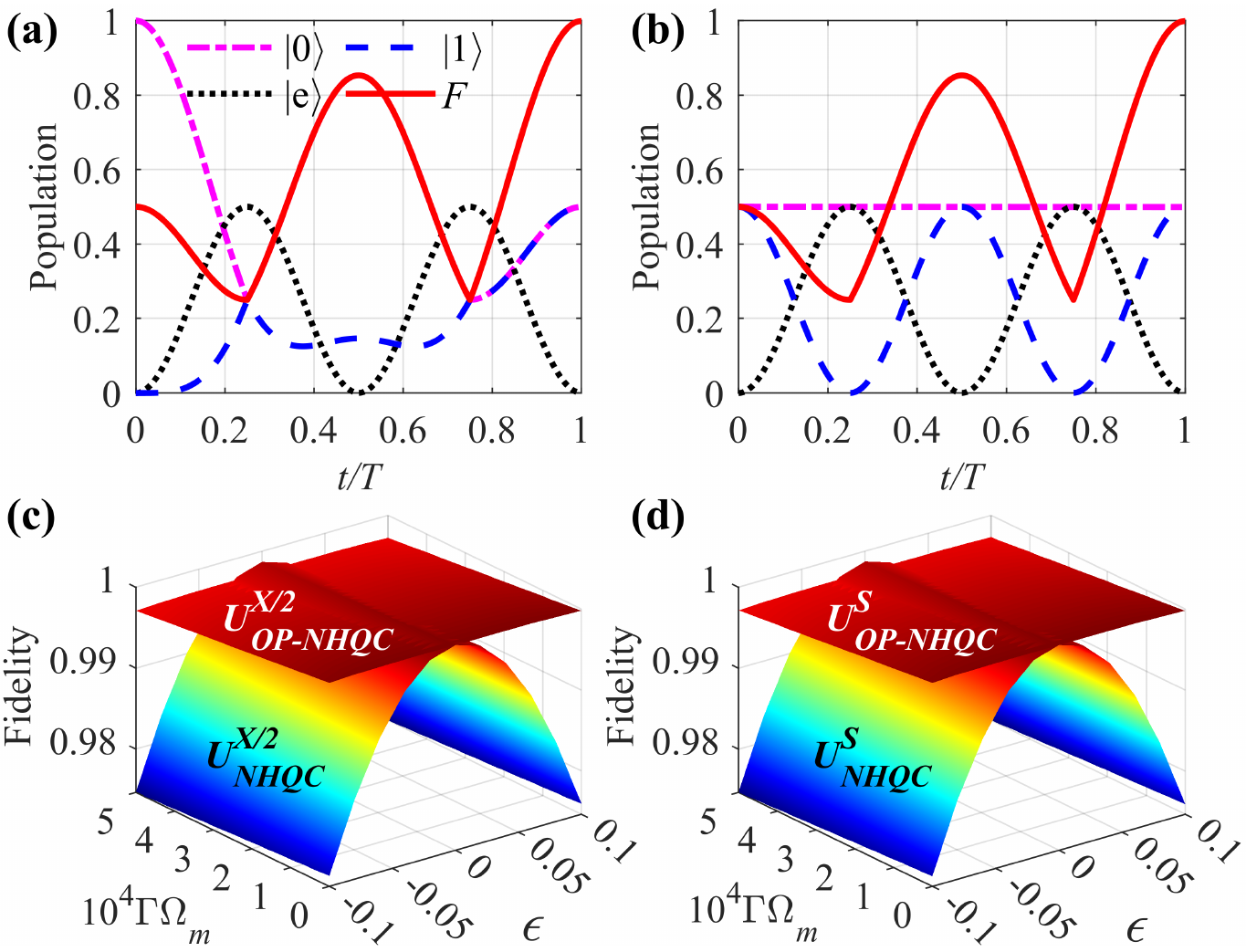}
		\caption{The state population and fidelity dynamics of (a) $X/2$ and (b) $S$  gates for our OP-NHQC scheme. The decoherence rates are chosen as a  moderate value $\Gamma= 2 \times 10^{-4}$. A comparison of the robustness of the (c) $X/2$ gate and (d) $S$ gate in the conventional NHQC and OP-NHQC schemes, considering both the control error $\epsilon$ and the decoherence effect with a uniform decoherence rate $\Gamma/\Omega_{m}$.}\label{fig4}
	\end{center}
\end{figure}

\subsection{The gate performance}

The gate performance is evaluated by using  the Lindblad master equation
\begin{equation}
	\begin{aligned}
		\dot\rho_{0} =& i\left[\rho_{0}, \mathcal{H}\right]  + \frac{1}{2}  \sum^4_{j=1} \Gamma_j \mathcal{L}(\sigma_j),
	\end{aligned}
	\label{master}
\end{equation}
where $\rho_{0}$ is the density matrix of the considered system and $\mathcal{L}(\sigma_j)=2\sigma_j\rho_{0} \sigma_j^\dagger-\sigma_j^\dagger \sigma_j \rho_{0} -\rho_{0} \sigma_j^\dagger \sigma_j$ represents the Lindbladian of the operator $\sigma_j$ with $\sigma_1=|e\rangle\langle 0|$, $\sigma_2=|e\rangle\langle 1|$, $\sigma_3=(|0\rangle\langle 0|-|e\rangle\langle e|)/ 2$,  $\sigma_4=(|1\rangle\langle 1|-|e\rangle\langle e|)/ 2$, and $\Gamma_j$ are the corresponding decoherence rates. When considering the current state-of-the-art technologies, the ratio of $\Gamma_j/\Omega_{m}$ can be as low as $\sim10^{-4}$, where $\Omega_{m}$ denotes the maximum value of $\Omega\left(t\right)$. For simplicity, we set $\Omega(t)=\Omega_{m} = 1$ and $\phi_{0}^{\prime} = 0$ to simulate the performance of our dynamically optimized NHQC scheme with and without considering the control error $\epsilon$.

We first verify the state population and state fidelity without considering the control error $\epsilon$. Here, we give two gates as examples, namely  $X/2$ and $S$ gates, whose parameters are $\{\theta = \pi/2, \gamma_g = \pi/2,  \phi=0\}$, and $\{\theta = 0, \gamma_g = \pi/2, \phi=0\}$, respectively. The initial states are set to be $|\psi_i\rangle_{X/2}  = |0\rangle$  and  $|\psi_i\rangle_S  = (|0\rangle+|1\rangle)/\sqrt{2}$. The state fidelity is defined  as $F_{X/2(S)} =_{X/2(S)}\langle \psi_f|\rho_0|\psi_f\rangle_{X/2(S)}$ to evaluate these two holonomic gates. And the corresponding ideal target states are $|\psi_f\rangle_{X/2} = (|0\rangle+|1\rangle)/2+i (|0\rangle-|1\rangle)/2$ and  $|\psi_f\rangle_S = (|0\rangle+i|1\rangle)/\sqrt{2}$. Here, the decoherence rates are chosen as a  moderate value $\Gamma_{1}=\Gamma_{2}=\Gamma_{3}=\Gamma_{4}=\Gamma= 2 \times 10^{-4} $.  As shown in Figs.~\ref{fig4}(a) and~\ref{fig4}(b), we obtain the state fidelities for $X/2$ and $S$ gate as $F_{X/2}=99.90\%$ and $F_S=99.89\%$, respectively. Then, to fully evaluate the average gate performance, we consider using six typical initial states as $\{|0\rangle, |1\rangle, (|0\rangle-|1\rangle)/\sqrt{2}, (|0\rangle+|1\rangle)/\sqrt{2}, (|0\rangle-i|1\rangle)/\sqrt{2}, (|0\rangle+i|1\rangle)/\sqrt{2}\}$, and the corresponding gate fidelity is defined as \cite{Julian_1960,IDIvonovic_1981,Klappenecker_2005}
\begin{equation}
	F^G_1=\frac{1}{6}{\sum^{6}_{k = 1}} {_k\langle\psi_i|U^\dagger \rho_{0} U|\psi_i\rangle_k}.
\end{equation}
It can be found that the average fidelities for the two gates are $F^G _{X/2}= 99.89\%$ and $F^G _S= 99.90\%$, respectively. Furthermore, the numerical results confirm that  when $\Gamma < 2 \times 10^{-4}$, the fidelities for both gates can exceed $99.90\%$. Then, we demonstrate the robustness of our dynamically optimized NHQC scheme considering both the control error $\epsilon$ and decoherence effect. As shown in Fig.~\ref{fig4}(c) and~\ref{fig4}(d), the decoherence rate is with $\Gamma_{j}=\Gamma \in \left[0,5\right] \times 10^{-4}$ and the error is within the range of $-0.1\leq\epsilon\leq0.1$. It is clear that our dynamically optimized NHQC scheme can substantially improve the gate robustness for the $X/2$ and $S$ gates, compared to the conventional NHQC scheme. This result indicates superior gate robustness of our protocol.

\section{Physical implementation}

As mentioned above, our dynamically optimized NHQC scheme has been shown to be an effective method to suppress the control error. In the following section, we will demonstrate the compatibility of our dynamically optimized NHQC scheme with the DFS encoding technique \cite{ZanardiP_1997,DuanLu-Ming_1997,LidarDA_1998}. This compatibility allows for the synchronous suppression of both the control error $\epsilon$ and the collective dephasing error $\delta$. These suppression effects are achieved by the use of superconducting quantum circuits.

\subsection{Single-logical-qubit holonomic gate}

In this section, we first show how to use the  capacitively coupled transmon qubits to encode a logical qubit within the DFS technology. As shown in Fig~\ref{fig1}(b), the Hamiltonian between two coupled transmon qubits can be expressed as
\begin{equation}
	\begin{aligned}
		\mathcal{H}_{ij}^c= & \sum_{l=i, j} \sum_{k=0}^\infty\left[k \omega_l-k(k-1) \alpha_l/2\right]|k\rangle_l\langle k| +\left\{g_{i j} \prod_{l=i, j}\left( |0\rangle_{l}\langle 1|+\sqrt{2}|1\rangle_{l}\langle2|\right)+\text { H.c. }\right\},
	\end{aligned}
\end{equation}
where $g_{ij}$ represents the coupling strength, and $\alpha_{i(j)}$ is the anharmonicity of transmon qubit $Q_{i(j)}$. By driving the transmon qubit $Q_{j}$ with an additional qubit frequency in the form of $\omega_{j}\left(t\right)=\omega_{j}+\varepsilon_{j}\sin\left(\nu_{j}t+\eta_{j}\right)$, we can then realize the periodically controllable manipulation between two neighboring transmons. By applying the Jacobi-Anger identity 
and moving the Hamiltonian into the interaction picture, we can then express the transformed Hamiltonian as
\begin{equation}
	\begin{aligned}
		\mathcal{H}_{i j}= & \sum_{n=-\infty}^{+\infty}(-i)^{n} J_{n}\left(\beta_j\right) g_{i j}\left\{|01\rangle_{i j}\langle 10| e^{i \Delta_{j} t} e^{-i n\left(\nu_{j} t+\eta_{j}\right)}\right. \\
		& +\sqrt{2}|02\rangle_{i j}\langle 11| e^{i\left(\Delta_{j}+\alpha_{j}\right) t} e^{-i n\left(\nu_{j} t+\eta_{j}\right)} \\
		& \left.+\sqrt{2}|11\rangle_{i j}\langle 20| e^{i\left(\Delta_{j}-\alpha_{i}\right) t} e^{-i n\left(\nu_{j} t+\eta_{j}\right)}\right\}+ \textrm{ H.c. }
	\end{aligned}
	\label{Eq:TransHamiltonian}
\end{equation}
where $\beta_{j}=\varepsilon_{j}/\nu_{j}$ and $\Delta_{j}=\omega_{j}-\omega_{i}$ is the frequency difference between two qubits. $J_{n}\left(\beta_{j}\right)$ represents the $n$-th Bessel function of the first kind. It is possible to achieve the single- or two-excitation subspaces by modulating the qubit-frequency driving parameters $\varepsilon_{j}$ and $\nu_{j}$, and the energy level diagram is illustrated in Fig.~\ref{fig1}(c). Below, we propose the utilization of two transmon qubits, denoted as $Q_{1}$ and $Q_{2}$, as a logical unit for the encoding of a single logical qubit (see Fig.~\ref{fig1}(b)). Each of these qubits is coupled with the auxiliary transmon qubit, designated as $Q_{a}$. In this way, the logical-qubit states are defined as $\mathcal{S}_{0}= $ Span$\{|0\rangle_{L} = |01\rangle_{12}$, $|1\rangle_{L} = |10\rangle_{12}\}$. The corresponding control Hamiltonian considering both the auxiliary and logical transmon qubits can be described
\begin{equation}
	\begin{aligned}
		\mathcal{H}_{1}= H_{a1}+H_{a2},
	\end{aligned}
	\label{Eq:H1}
\end{equation}
where $H_{al}$ $(l=1,2)$ represents the interaction Hamiltonian between qubits $Q_{a}$ and $Q_{l}$. By setting the driving parameters to satisfy $\Delta_{1}=\nu_{1}$ and $\Delta_{2}=\nu_{2}$, the effective resonant interaction Hamiltonian in the single-excitation subspace of $H_{1}$ can be written as
\begin{equation}
	\begin{aligned}
		\mathcal{H}^{1}_{L}&= g_{a1}^{\prime} e^{-i\eta_{1}} |01\rangle_{a1}\langle 10|+g_{a2}^{\prime} e^{-i\eta_{2}} |01\rangle_{a2} \langle 10|+\textrm{H.c.}\\&
		=ge^{-i\eta_{1}}|0\rangle_{a}\langle1|\otimes\left(\cos\frac{\theta}{2}|1\rangle_{1}\langle 0|-\sin\frac{\theta}{2}e^{-i\eta}|1\rangle_{2} \langle 0|\right)\quad+\textrm{H.c.}
	\end{aligned}
	\label{Eq:H1L}
\end{equation}
where $g_{al}^{\prime}=J_{1}\left(\beta_{l}\right)g_{al}$ and $\eta_{l}^{\prime}=\eta_{l}+\pi/2$. We define $g=\sqrt{g_{a1}^{\prime2}+g_{a2}^{\prime2}}$, $g_{a1}^{\prime}=g\cos\frac{\chi}{2}$, $g_{a2}^{\prime}=g\sin\frac{\chi}{2}$, and $\eta=\eta_{2}^{\prime}-\eta_{1}^{\prime}+\pi$. Alternatively, the Hamiltonian $\mathcal{H}^{1}_{L}$ considering the three transmon qubits can be rewritten as
\begin{equation}
	\begin{aligned}
		\mathcal{H}^{1}_{L}
		&=ge^{-i\eta_{1}}\left[ \left(\cos\frac{\chi}{2}|010\rangle_{a12}\langle 100|-\sin\frac{\chi}{2}e^{-i\eta}|001\rangle_{a12} \langle 100|\right)\right. \\&\left.+\left(\cos\frac{\chi}{2}|011\rangle_{a12}\langle 101|-\sin\frac{\chi}{2}e^{-i\eta}|011\rangle_{a12} \langle 110|\right) \right]+\textrm{H.c.}\\&
		=g\left[e^{-i\eta_{1}}|B_{1}\rangle_{L^{\prime}}\langle E_{1}|+e^{i\eta_{1}}|E_{2}\rangle_{L^{\prime\prime}}\langle B_{2}|\right]+\textrm{H.c.},
	\end{aligned}
	\label{Eq:H1LUpdate}
\end{equation}
in which $|B_{1}\rangle_{L^{\prime}}=\cos\frac{\chi}{2}|010\rangle_{a12}-\sin\frac{\chi}{2}e^{-i\eta}|001\rangle_{a12}$ and $|B_{2}\rangle_{L^{\prime\prime}}=\cos\frac{\chi}{2}|101\rangle_{a12}-\sin\frac{\chi}{2}e^{-i\eta}|110\rangle_{a12}$ represent the dressed states between two types of different DFS subspaces; and $|E_{1}\rangle_{L^{\prime}}=|100\rangle_{a12}$ and $|E_{2}\rangle_{L^{\prime\prime}}=|011\rangle_{a12}$ are the ancillary states. Under these settings, two three-dimensional DFS  $\mathcal{S}_{1}= $ Span$\{|0\rangle_{L^{\prime}}=|001\rangle_{a12}, |1\rangle_{L^{\prime}}=|010\rangle_{a12}, |E_{1}\rangle_{L^{\prime}}=|100\rangle_{a12}\}$ and $\mathcal{S}_{2}= $ Span$\{|0\rangle_{L^{\prime\prime}}=|101\rangle_{a12}, |1\rangle_{L^{\prime\prime}}=|110\rangle_{a12}, |E_{2}\rangle_{L^{\prime\prime}}=|011\rangle_{a12}\}$ are identified.
There are two block-diagonal three-level structures that can be used to construct holonomic quantum gates, which can suppresses both the control error $\epsilon$ and collective dephasing error $\delta$.

\begin{figure}[tbp]
	\begin{center}
		\includegraphics[width=0.65\columnwidth]{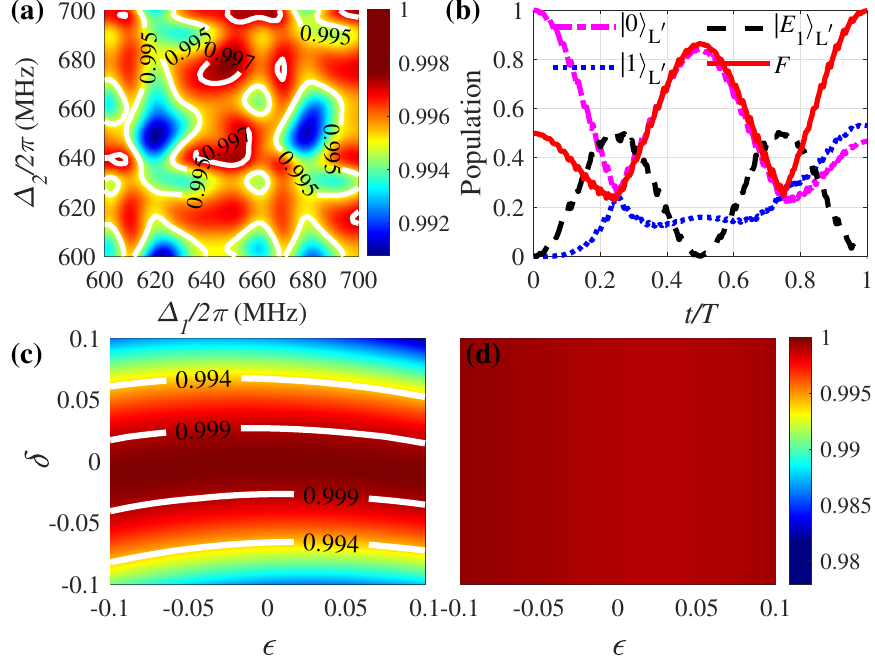}
		\caption{(a) The gate fidelity as a function of the qubit frequency differences $\Delta_{1}$ and $\Delta_{2}$ for the $X/2$ gate. (b) State populations and the state fidelity dynamics of the $X/2$ gate. Gate robustness of the $X/2$ gate in OP-NHQC (c) without and (d) with the DFS encoding, considering both the control error $\epsilon$ and the collective dephasing error $\delta$.}\label{fig6}
	\end{center}
\end{figure}

Moreover, the Hamiltonian in Eq.~\eqref{Eq:H1LUpdate} exhibits a structural similarity to that presented in Eq.~\eqref{Eq:HI}. It is therefore possible to utilize the subspace $\mathcal{S}_{1}$ to encode our dynamically optimized NHQC scheme in a manner that simultaneously suppresses the control error $\epsilon$ and collective dephasing error $\delta$.  
In consequence, we divide our dynamically optimized NHQC scheme into four parts, with the corresponding control parameters expressed as follows:
\begin{equation}
	\begin{aligned}
		t&\in\left[0,T_{1}\right],  \ \ \quad gT_{1}= \frac{\pi}{2}, \quad\quad\quad\quad \eta_{1}=\eta_{0}^{\prime},\\
		t&\in\left[T_{1},T_{2}\right],  \quad g\left(T_{2}-T_{1}\right)= \frac{\pi}{2}, \quad \eta_{1}=\eta_{0}^{\prime}+\pi-\frac{\gamma_{g}}{2},\\
		t&\in\left[T_{2},T_{3}\right],  \quad g\left(T_{3}-T_{2}\right)= \frac{\pi}{2}, \quad \eta_{1}=\eta_{0}^{\prime}+\pi-\gamma_{g},\\
		t&\in\left[T_{3},T\right],  \quad g\left(T-T_{3}\right)= \frac{\pi}{2}, \quad \ \ \eta_{1}=\eta_{0}^{\prime}+2\pi-\frac{3\gamma_{g}}{2}.
	\end{aligned}
	\label{Eq:OptiNHQC_DFS}
\end{equation}
After substituting the above parameters into Eq.~\eqref{Eq:H1LUpdate}, the evolution operator can be written as
\begin{equation}
	\begin{aligned}
		U_{\rm{L}}\left(0,T\right)=|0\rangle_{a}\langle0|\otimes U_{\rm{L}}^{\prime}+|1\rangle_{a}\langle1|\otimes U_{\rm{L}}^{\prime\prime},
	\end{aligned}
\end{equation}
where $U_{\rm{L}}^{\prime}=e^{i\frac{\gamma_{g}}{2}}e^{-i\frac{\gamma_{g}}{2}\mathbf{n^{\prime}}\cdot \mathbf{\sigma^{\prime}}}$ and $U_{\rm{L}}^{\prime\prime}=e^{-i\frac{\gamma_{g}}{2}}e^{-i\frac{\gamma_{g}}{2}\mathbf{n^{\prime\prime}}\cdot \mathbf{\sigma^{\prime}}}$. The unit vector are $\mathbf{n^{\prime}}=\left(\sin\chi\cos\eta,\sin\chi\sin\eta,\cos\chi\right)$ and $\mathbf{n^{\prime\prime}}=\left(-\sin\chi\cos\eta,-\sin\chi\sin\eta,\cos\chi\right)$. $\mathbf{\sigma^{\prime}}=\left(\sigma^{\prime}_{x},\sigma^{\prime}_{y},\sigma^{\prime}_{z}\right)$, the Pauli matrices are defined as $\sigma^{\prime}_{x}=|0\rangle_{L}\langle1|+|1\rangle_{L}\langle0|$, $\sigma^{\prime}_{y}=-i|0\rangle_{L}\langle1|+i|1\rangle_{L}\langle0|$ and $\sigma^{\prime}_{z}=|0\rangle_{L}\langle0|-|1\rangle_{L}\langle1|$.

Nevertheless, the performance of the proposed holonomic gate is unavoidably constrained by the qubit decoherence effects inherent to the practical physical system. Here, we consider using the Lindblad master equation of
\begin{equation}
	\begin{aligned}
		&\dot{\rho}_{1}=-i\left[\mathcal{H}_S, \rho_1\right]  +\sum_{l=a, 1,2}\sum_{k=1}^2\left\{\frac{\kappa_{-}^l}{2} \mathcal{L}\left(\sqrt{k}|k-1\rangle_{l}\langle k|\right)+\frac{\kappa_z^l}{2} \mathcal{L}\left(k |k\rangle_{l}\langle k|\right)\right\},
	\end{aligned}
\end{equation}
to numerically simulate the decoherence effects and the high-order oscillating terms of the system, where $\kappa_{-}^{l}$ and $\kappa_{z}^{l}$ represent the relaxation and dephasing rates of the $l$-th transmon, respectively. Here, we consider the $X/2$ gate for a typical example to show the gate fidelity for our dynamically optimized NHQC scheme. For a given general initial evolution state $|\psi_{1}\rangle=\cos\theta_{1}|0\rangle_{L^{\prime}}+\sin\theta_{1}|1\rangle_{L^{\prime}}$, the ideal final state is calculated as $|\psi_{f_{m}}\rangle=U^{\prime}_{L}|\psi_{1}\rangle$. The gate fidelity can then be defined as $F^{G}_{m}=\frac{1}{2\pi}\int_{0}^{2\pi}\langle\psi_{f_{m}}|\rho_{1}|\psi_{f_{m}}\rangle d\theta_{1}$. Take into account the current state-of-the-art experiments, the decoherence parameters are set as $\kappa=\kappa_{-}^{a}=\kappa_{z}^{a}=\kappa_{-}^{1}=\kappa_{z}^{1}=\kappa_{-}^{2}=\kappa_{z}^{2}= 2\pi \times 2$ kHz for each physical transmon qubit. Here, the anharmonicity for each transmon qubit is $\alpha_{a}=2\pi\times 210$ MHz, $\alpha_{1}=2\pi\times 273$ MHz, and $\alpha_{2}=2\pi\times 230$ MHz, respectively. The coupled strength is set as $g_{a1}=g_{a2}=2\pi\times 18.2$ MHz. For the $X/2$ gate, the qubit-driving parameters of the two pairs of parametrically coupled transmons are equal, i.e., $J_{1}\left(\beta_{2}\right)/J_{2}\left(\beta_{2}\right)=1$. As shown in Fig.~\ref{fig6}(a), the numerical simulation result shows that the qubits frequency difference $\Delta_{1}$ and $\Delta_{2}$ can affect the gate fidelity. When choosing $\Delta_{1}=2\pi\times 649$ MHz and $\Delta_{2}=2\pi\times 679$ MHz, the gate fidelity of the $X/2$ gate can reach at $99.77\%$. We also define the state fidelity as $F_{m}=\langle\psi_{f_{m}}|\rho_{1}|\psi_{f_{m}}\rangle$ to evaluate the gate's population. Here, the initial state of the system is defined as $|\psi_{1}\rangle=|0\rangle_{L^{\prime}}$, and the ideal final state is thus $|\psi_{f_{X/2}}\rangle=\left[\left(1+i\right)|0\rangle_{L^{\prime}}+\left(1-i\right)|1\rangle_{L^{\prime}}\right]$. As shown in Fig.~\ref{fig6}(b), the state fidelity can reach $99.74\%$. Then, we further evaluate gate robustness of our dynamically optimized NHQC scheme with the DFS encoding. Considering the control error $\epsilon$ and collective dephasing error $\delta$, the Hamiltonian without DFS encoding in Eq.~\eqref{Eq:HI} is written as $\mathcal{H}^E(t) =  (1+\epsilon)\mathcal{H}(t) + \delta \Omega_m|e\rangle\langle e|$. For comparison, the corresponding one with DFS encoding in Eq.~\eqref{Eq:H1L} is $\mathcal{H}^E_L(t) =  (1+\epsilon)\mathcal{H}^1_L+ \delta g (|0\rangle_{L^{\prime}}\langle0|+|1\rangle_{L^{\prime}}\langle1|+|E_1\rangle_{L^{\prime}}\langle E_1|)$. As shown in Figs.~\ref{fig6}(c) and~\ref{fig6}(d), combining our dynamically optimized NHQC scheme and the DFS encoding technique, it can greatly suppress the control error $\epsilon$ and collective dephasing error $\delta$ at the same time without considering decoherence rate.

\subsection{Two-logical-qubit holonomic gate}

\begin{figure}[tbp]
	\begin{center}
		\includegraphics[width=0.45\columnwidth]{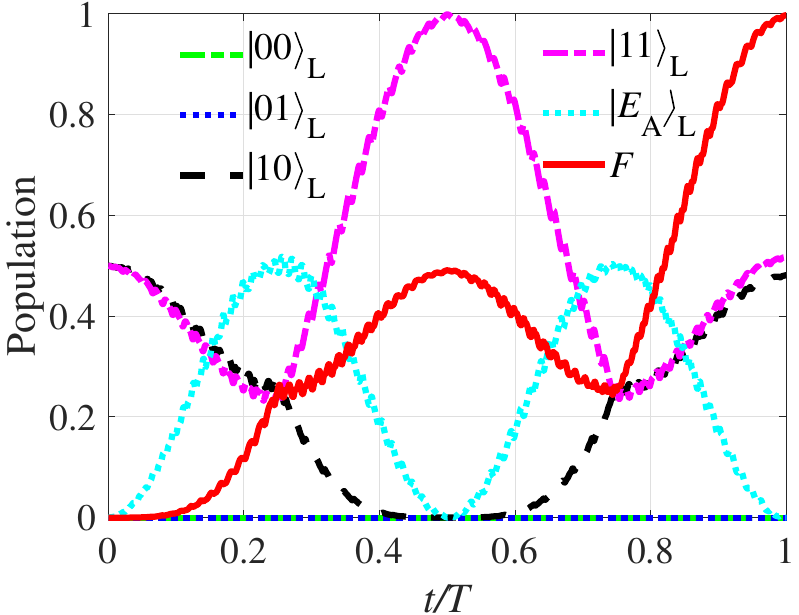}
		\caption{State dynamics of the two-qubit gate $U_{T}^{\prime}$. The initial state is set to be $(|10\rangle_L-i|11\rangle_L)/\sqrt{2}$, and the decoherence rate is  $\kappa = 2\pi \times 2$ kHz. The state fidelity is $99.72\%$.}\label{fig7}
	\end{center}
\end{figure}

Here, we turn to introduce how to realize non-trivial two-qubit gate with the DFS encoding.  The physical qubits are connected as shown in Fig.~\ref{fig1}(b), where we use the qubits $Q_1$, $Q_2$, $Q_{3}$ and $Q_4$ to realize the  two-qubit logical subspace. The corresponding Hamiltonian is written as
\begin{equation}
	\begin{aligned}
		\mathcal{H}_{2}= H_{13}+H_{14}.
	\end{aligned}
	\label{Eq:H2}
\end{equation}
Similar to the single-qubit DFS encoding, there is a four-dimensional DFS $\mathcal{S}_{3}=\left\{|00\rangle_{L}, |01\rangle_{L},|10\rangle_{L}, |11\rangle_{L}\right\}$. Specifically, the basis vectors of the two-qubit logical gates are defined as
\begin{equation}\label{Eq:Two-qubit_DFS2}
	\begin{aligned}
		\mathcal{S}_{3}= \mathrm{Span} \{&|00\rangle_{L}=|0101\rangle_{1234}, |01\rangle_{L}=|0110\rangle_{1234}, \notag \\
		&|10\rangle_{L}=|1001\rangle_{1234},|11\rangle_{L}=|1010\rangle_{1234}\}. \notag\\
	\end{aligned}
\end{equation}
Therefore, under this four-dimensional two-qubit DFS subspace $\mathcal{S}_{3}$, the Hamiltonian of the coupled system in Eq.~\eqref{Eq:H2} changes to
\begin{equation}
	\begin{aligned}
		\mathcal{H}^2_L(t) &= g_{13}^{\prime} e^{i\left(\Delta_{3}+\alpha_{1}\right)t} e^{-i\left(\nu_{3}+\eta_{3}^{\prime}\right)} | 11 \rangle_{13}\langle 20|+g_{14}^{\prime} e^{i\left(\Delta_{4}+\alpha_{1}\right)t} e^{-i\left(\nu_{4}+\eta_{4}^{\prime}\right)} |11 \rangle_{14}\langle 20|+ \textrm{H.c.},
	\end{aligned}
	\label{Eq:H2L}
\end{equation}
where $g_{13}^{\prime}=\sqrt{2}g_{13}J_{1}\left(\beta_{3}\right)$, $g_{14}^{\prime}=\sqrt{2}g_{14}J_{1}\left(\beta_{4}\right)$, $\eta_{3}^{\prime}=\eta_{3}+\pi/2$ and $\eta_{4}^{\prime}=\eta_{4}+\pi/2$. By modulating the qubit-frequency driving parameters $\nu_{3}=\Delta_{3}+\alpha_{1}$, $\nu_{4}=\Delta_{4}+\alpha_{1}$ and setting $g_{13}^{\prime}=g^{\prime}\cos\frac{\vartheta}{2}$, $g_{14}^{\prime}=g^{\prime}\sin\frac{\vartheta}{2}$, in the two-qubit logical subspace Span$\{ |10\rangle_{L}, |11\rangle_{L}, |E_{A}\rangle_{L}\}$, the resonant interaction Hamiltonian can be reduced as
\begin{equation}
	\begin{aligned}
		\mathcal{H}^3_L(t) &= g^{\prime}\left(\cos\frac{\vartheta}{2} e^{-i\eta_{3}^{\prime}} |11 \rangle_{13}\langle 20|+\sin\frac{\vartheta}{2} e^{-i\eta_{4}^{\prime}} |11 \rangle_{14}\langle 20|\right)+ \textrm{H.c.},\\&
		=g^{\prime}e^{-i\eta_{3}^{\prime}}\left(\cos\frac{\vartheta}{2} |11 \rangle_{L}-\sin\frac{\vartheta}{2} e^{-i\eta^{\prime}} |10 \rangle_{L}\right) {_{L}\langle E_{A}|}+ \textrm{H.c.},
	\end{aligned}
	\label{Eq:H2Lx}
\end{equation}
where $\eta^{\prime}=\eta_{4}^{\prime}-\eta_{3}^{\prime} +\pi$, and $|E_{A}\rangle_{L}=|0200\rangle_{1234}$ is regard as an ancillary state. Following the same procedure in Eq.~\eqref{Eq:OptiNHQC_DFS},
the final evolution operator in the two-qubit logical subspace Span$\{ |10\rangle_{L}, |11\rangle_{L}, |E_{A}\rangle_{L}\}$ can be expressed as
\begin{equation}
	\begin{aligned}
		U_{\rm{T}}^{\prime}\left(\vartheta,\eta^{\prime},\gamma_{g}\right)
		=e^{i\frac{\gamma_{g}}{2}}\left(
		\begin{array}{ccc}
			\upsilon & -i\beta^{*} & 0 \\
			-i\beta & \upsilon^{*} & 0\\
			0      & 0          & e^{-i\frac{3\gamma_{g}}{2}}\\
		\end{array}
		\right),
	\end{aligned}
	\label{Eq:TwoQubit_DFS}
\end{equation}
where $\upsilon$ and $\beta$ satisfy $\upsilon\equiv \cos\left(\gamma_{g}/2\right)-i\cos\vartheta \sin\left(\gamma_{g}/2\right)$, and $\beta\equiv \sin\vartheta \sin\left(\gamma_{g}/2\right)e^{i\eta^{\prime}}$. Then, we can use the operator $U_{T}^{\prime}\left(\vartheta,\eta^{\prime},\gamma_{g}\right)$ to realize non-trivial two-qubit gates.
Thus, within the logical-qubit subspace Span$\{|00\rangle_{L}$, $|01\rangle_{L}$, $|10\rangle_{L}$, $|11\rangle_{L}\}$, by setting $\chi=\pi/2$, $\eta=0$, and $\gamma_{g}=\pi/2$, we can get a non-trivial two-qubit CNOT gate \cite{DMZajac_2018,NohTaewan_2018,XieTianyu_2023}
\begin{equation}
	\begin{aligned}
		U_{\rm{T}}^{\prime}\left(\pi/2,0,\pi/2\right)=&\left(
		\begin{array}{cccc}
			1 & 0 & 0 & 0  \\
			0 & 1 & 0 & 0  \\
			0 & 0 & 0 & 1 \\
			0 & 0 & 1 & 0  \\
		\end{array}
		\right).
	\end{aligned}
	\label{Eq:entangle_Cnot}
\end{equation}
Here, the qubit-decoherence rates are also set as $\kappa=\kappa_{-}^{1}=\kappa_{z}^{1}=\kappa_{-}^{2}=\kappa_{z}^{2} =\kappa_{-}^{3}=\kappa_{z}^{3}=\kappa_{-}^{4}=\kappa_{z}^{4}=2\pi \times 2$ kHz. The other qubit parameters are set as $\alpha_{3}=2\pi\times 216$ MHz, $\alpha_{4}=2\pi\times 224$ MHz, $\Delta_{3}=2\pi\times 663$ MHz, $\Delta_{4}=2\pi\times 693$ MHz, $g_{13}=g_{14}=2\pi\times 13$ MHz and $J_{1}\left(\beta_{3}\right)/J_{1}\left(\beta_{4}\right)=1$. Under this setting, the two-qubit gate fidelity can be as high as $99.69\% $. As shown in Fig.~\ref{fig7}, we consider the initial state as $(|10 \rangle_{L}-i|11 \rangle_{L})/\sqrt{2}$, the state fidelity can reach $99.72\% $. It is noteworthy that our two-qubit holonomic gates with DFS encoding are also capable of effectively resisting the $\delta$ error caused by collective dephasing.

\section{Conclusion}
In conclusion, we propose a dynamically optimized NHQC scheme to enhance the performance of holonomic gates. Leveraging the dynamically corrected gate technique, our scheme can suppress control errors up to the fourth order. Notably, by integrating the DFS encoding in superconducting quantum circuits, the proposed approach becomes robust against both control error and collective dephasing error. Compared to previous implementations, numerical simulations demonstrate that our dynamically optimized NHQC strategy achieves outstanding performance. Thus, our scheme holds great promise for the future development of scalable and fault-tolerant holonomic quantum computation.

\data{The data that support the findings of this study are available on request from the corresponding author upon reasonable request.}

\ack{This work was supported by the National Natural Science Foundation of China (Grant No. 11905065, 62171144, 12305019), the Guangxi Science Foundation (Grant No. AD22035186, 2021GXNSFAA220011) and the Project of Improving the Basic Scientific Research Ability of Young and Middle-aged Teachers in Universities of Guangxi (Grant No. 2023KY0815).}

\section*{Appendix}
\appendix

\section{The composite NHQC scheme}\label{AP1}

\begin{figure*}[htbp]
	\begin{center}
		\includegraphics[width=0.95\columnwidth]{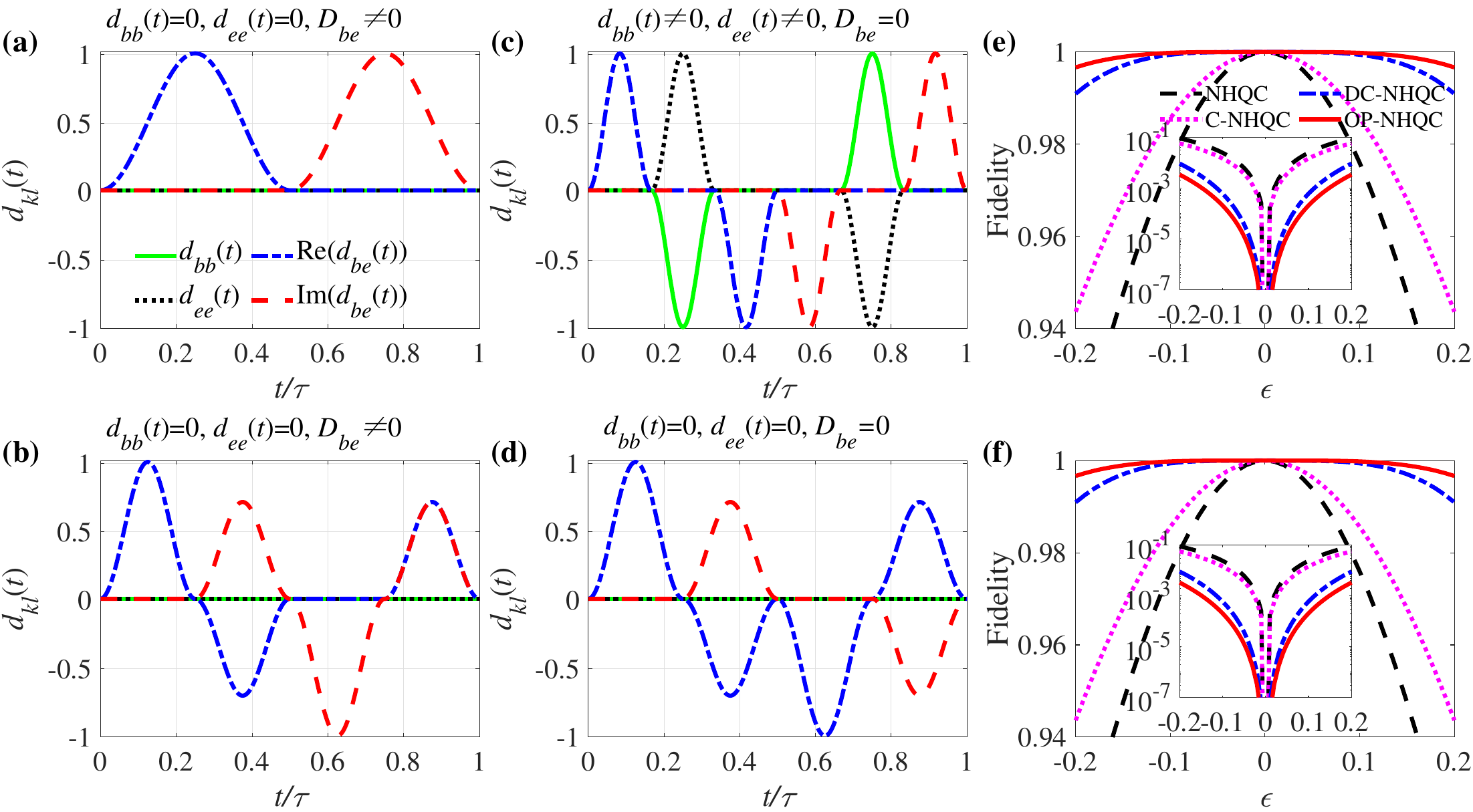}
		\caption{The dynamical phase accumulated over time evolution for (a) the conventional NHQC scheme, (b) composite NHQC scheme, (c) dynamically corrected NHQC scheme and (d) our dynamically optimized NHQC scheme, with the holonomic $X/2$ gate taken as an example. Robustness of the OP-NHQC scheme compared to the NHQC, C-NHQC, DC-NHQC schemes, where (e) denotes $X/2$ and (f) the $S$  gate.}\label{FigDPhase}
	\end{center}
\end{figure*}

Below we will introduce how to use the composite NHQC (C-NHQC) scheme \cite{ZhuZhennan_2019,XuGuoFu_2017} to construct the holonomic gates. Based on the composite pulse technology, the conventional NHQC scheme in Eq.~\eqref{Eq:Con_NHQC_Sche} can be extend to the composite holonomic gate form. The evolution operation has the relation: $\left[U\left(\theta,\phi,\gamma_{g}/N\right)\right]^{N}=U\left(\theta,\phi,\gamma_{g}\right)$, $N$ is the number of composite pulse. Here we just consider the case of $N=2$. The control parameters of composite NHQC scheme are given by 
\begin{equation}
	\begin{aligned}
		t\in\left[0,\tau_{1}\right],  \int_{0}^{\tau_{1}}\Omega(t)dt&= \frac{\pi}{2}, \quad \phi_{0}=\phi_{0}^{\prime},\\
		t\in\left[\tau_{1},\tau_{2}\right],  \int_{\tau_{1}}^{\tau_{2}}\Omega(t)dt&= \frac{\pi}{2}, \quad \phi_{0}=\phi_{0}^{\prime}+\pi-\frac{\gamma_{g}}{2},\\
		t\in\left[\tau_{2},\tau_{3}\right],  \int_{\tau_{2}}^{\tau_{3}}\Omega(t)dt&= \frac{\pi}{2}, \quad \phi_{0}=\phi_{0}^{\prime},\\
		t\in\left[\tau_{3},\tau\right],  \int_{\tau_{3}}^{\tau}\Omega(t)dt&= \frac{\pi}{2}, \quad \phi_{0}=\phi_{0}^{\prime}+\pi-\frac{\gamma_{g}}{2}.
	\end{aligned}
	\label{Eq:TwoLoop_NHQC_Sche}
\end{equation}
The final evolution has the same form as shown in Eq.~\eqref{Eq:UU}. When the holonomic gates suffer from the influence of control error $\epsilon$, the operation in the dressed basis $\{|b\rangle, |d\rangle\}$ can be expressed as
\begin{equation}
	\begin{aligned}
		U^{\epsilon}_{\rm{tl}} = & |d\rangle \langle d|+\left [\sin^2\mu\pi(\cos\frac{\gamma_{g}}{2}+e^{i\frac{\gamma_{g}}{2}})/2 +\sin^4\mu\pi e^{i\gamma_{g}}+ \cos^2\frac{\mu\pi}{2}(1-3\sin^2\frac{\mu\pi}{2}) \right ]|b\rangle \langle b|.
	\end{aligned}
	\label{Eq:UeTLNHQC}
\end{equation}
Furthermore, the gate fidelity can be expanded as
\begin{equation}
	\begin{aligned}
		F_{\rm{tl}} =&\frac{1}{2}\left| 1+ \frac{1}{2}\sin^{2}\mu\pi\left(\cos\frac{\gamma_{g}}{2}+e^{i\frac{\gamma_{g}}{2}}\right)e^{-i\gamma_{g}}+\sin^{4}\frac{\mu\pi}{2}+\cos^{2}\frac{\mu\pi}{2}\left(1-3\sin^{2}\frac{\mu\pi}{2}\right)e^{-i\gamma_{g}}  \right| \\
		\approx & 1-\epsilon^{2}\pi^{2}\sin^{2}\frac{\gamma_{g}}{4}\cos^{2}\frac{\gamma_{g}}{2}.
	\end{aligned}
	\label{Eq:FUeTLNHQC}
\end{equation}
This result demonstrates that, similarly to the conventional NHQC scheme, the composite NHQC is only capable of suppressing the control error to the second order. Meanwhile, as illustrated in Fig.~\ref{FigDPhase}(b), the dynamical phase element $d_{bb}(t)$ is exhibited to be zero at all times, in a manner analogous to that of the conventional NHQC scheme in Fig.~\ref{FigDPhase}(a). However, it is important to note that both the composite NHQC scheme and the conventional NHQC scheme are unable to eliminate the residual dynamical phase induced by the non-zero cross-coupling term $d_{be}(t)$, i.e., $D_{be}\neq0$, as shown in Figs.~\ref{FigDPhase}(a) and~\ref{FigDPhase}(b). Consequently, as depicted in Figs.~\ref{FigDPhase}(e) and~\ref{FigDPhase}(f), it is observed that this residual dynamical phase can directly influence the robustness of the holonomic gates.

\section{The dynamically corrected NHQC scheme}\label{AP2}

With the guidance of the dynamical correction method \cite{KhodjastehKaveh_2009,WangXin_2012,RongXing_2015}, the dynamically corrected NHQC (DC-NHQC) scheme \cite{LiSai_2021exper,LiSai_2021,LiuBaoJie_2021super} can further enhance the robustness to against the control error $\epsilon$ by inserting two pulses at the middle point in each half of the evolution path in the conventional NHQC. Accordingly, the control parameters of the Hamiltonian are given by
\begin{eqnarray}\label{Eq:Dyna_NHQC_Sche}
	t\in\left[0,\tau_{1}\right],  \int_{0}^{\tau_{1}}\Omega(t)dt&=& \frac{\pi}{4}, \quad \phi_{0}=\phi_{0}^{\prime},\notag \\
	t\in\left[\tau_{1},\tau_{2}\right],  \int_{\tau_{1}}^{\tau_{2}}\Omega(t)dt&=& \frac{\pi}{2}, \quad \phi_{0}=\phi_{0}^{\prime}+\frac{\pi}{2},\notag \\
	t\in\left[\tau_{3},\tau_{4}\right],  \int_{\tau_{3}}^{\tau_{4}}\Omega(t)dt&=& \frac{\pi}{4}, \quad \phi_{0}=\phi_{0}^{\prime},\notag \\
	t\in\left[\tau_{4},\tau_{5}\right],
	\int_{\tau_{4}}^{\tau_{5}}\Omega(t)dt&=& \frac{\pi}{4}, \quad \phi_{0}=\phi_{0}^{\prime}+\pi-\gamma_{g},\notag \\
	t\in\left[\tau_{5},\tau_{6}\right],  \int_{\tau_{5}}^{\tau_{6}}\Omega(t)dt&=& \frac{\pi}{2}, \quad \phi_{0}=\phi_{0}^{\prime}-\frac{\pi}{2}-\gamma_{g},\notag \\
	t\in\left[\tau_{6},\tau\right],  \int_{\tau_{6}}^{\tau}\Omega(t)dt&=& \frac{\pi}{4}, \quad \phi_{0}=\phi_{0}^{\prime}+\pi-\gamma_{g}.\notag\\
\end{eqnarray} 
Notably, the whole process above can also be used to realize universal single-qubit holonomic gates, as the same in Eq.~\eqref{Eq:UU}. When the control error $\epsilon$ is taken into account, the holonomic gates in the dressed basis $\{|b\rangle, |d\rangle\}$ will change to
\begin{equation}
	\begin{aligned}
		U^{\epsilon}_{\rm{dc}} = &|d\rangle \langle d|+ \left [\cos^4\frac{\mu\pi}{2}+ \left (\sin^2\frac{\mu\pi}{2}
		+\frac{1}{4}{\sin^2\mu\pi}\right)e^{i\gamma_g} \right ]|b\rangle \langle b|.
	\end{aligned}
	\label{Eq:UeDCNHQC}
\end{equation}
For the dynamically corrected NHQC scheme, the gate fidelity can be expanded as
\begin{equation}
	\begin{aligned}
		F_{\rm{dc}} =&\frac{1}{2}\left| 1+ \sin^2\frac{\mu\pi}{2} +\frac{1}{4}{\sin^2\mu\pi} +\cos^4\frac{\mu\pi}{2}e^{-i\gamma_g}  \right| \\
		\approx & 1-\epsilon^4\pi^4(1-\cos\gamma_g)/32.
	\end{aligned}
	\label{Eq:FUeDCNHQC}
\end{equation}
The calculated results indicate that the dynamically corrected NHQC scheme can suppress the control error up to the fourth order. As shown in Fig.~\ref{FigDPhase}(c), the residual dynamical phase resulting from the nonzero cross-coupling term $d_{be}\left(t\right)$ vanishes following the cyclic evolution, i.e., $D_{be}=0$, in contrast to both the conventional NHQC scheme and the composite NHQC scheme. This outcome indicates that the elimination of the residual dynamical phase is instrumental in enabling the dynamically corrected NHQC scheme to suppress the control error to the fourth order. Moreover, in contrast to the conventional NHQC scheme and the composite NHQC scheme, the dynamically corrected NHQC scheme cannot always preserve $d_{bb}\left(t\right)=0$, but only ensures the final cancellation of the integral of $d_{bb}(t)$. This discrepancy leads to a compromise in the gate robustness of the dynamically corrected NHQC scheme.

\begin{figure}[htbp]
	\begin{center}
		\includegraphics[width=0.75\columnwidth]{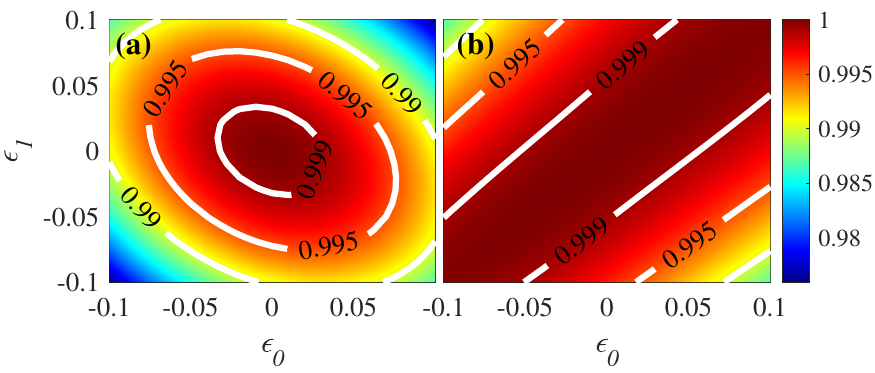}
		\caption{The gate fidelity as functions of control errors $\epsilon_{0}$ and $\epsilon_{1}$ for the $X/2$ gate in (a) conventional NHQC and (b) OP-NHQC scheme. One can find that although the control error between the magnitudes of $\Omega_0(t)$ and $\Omega_1(t)$ can not simultaneously change, i.e., $\Omega_{0}(t)\rightarrow (1+\epsilon_{0})\Omega_{0}(t)$ and $\Omega_{1}(t)\rightarrow (1+\epsilon_{1})\Omega_{1}(t)$, the OP-NHQC scheme still outperforms the conventional NHQC scheme.}\label{Fig_A1}
	\end{center}
\end{figure}

Fortunately, our dynamically optimized NHQC scheme not only eliminates the previously neglected cross-coupling terms but also avoids imposing additional geometric constraint, as is typically required in dynamically corrected NHQC schemes. As a result, our dynamically optimized NHQC scheme enhances the holonomic gate's resistance to the control error $\epsilon$ up to the fourth order and outperforms the dynamically corrected schemes. As demonstrated in Fig.~\ref{FigDPhase}(d), our dynamically optimized NHQC scheme ensures that $d_{bb}\left(t\right)=0$ throughout the evolution process and $D_{be}=0$ at the final time of cyclic evolution. Furthermore, as shown in in Figs.~\ref{FigDPhase}(e) and~\ref{FigDPhase}(f), a comparison of the gate-fidelity performance between different schemes verifies this claim.


\section{The robustness of OP-NHQC scheme against random fluctuation errors}\label{AP3}

\begin{figure}[tbp]
	\begin{center}
		\includegraphics[width=0.60\columnwidth]{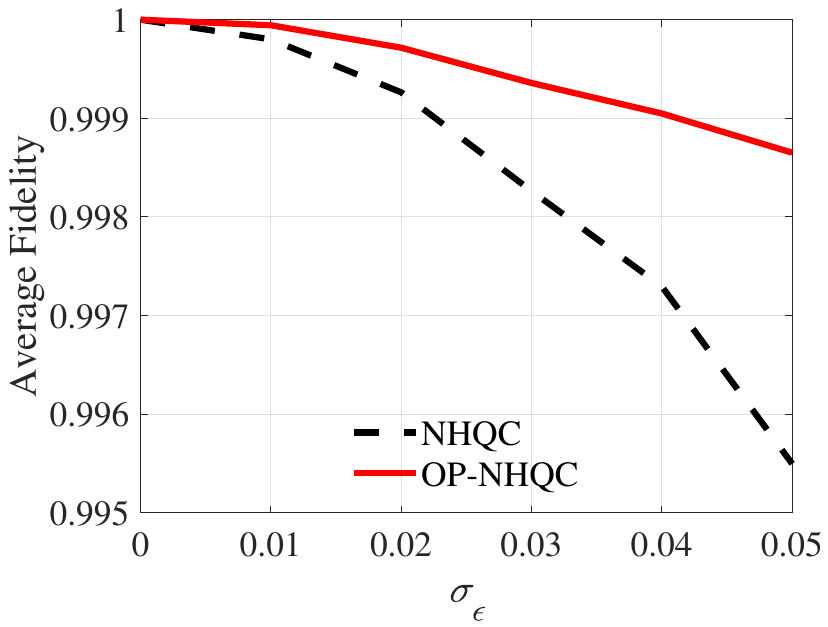}
		\caption{The robustness of $X/2$ gate against the static Gaussian noise. In the simulation, we consider the static noise model with Gaussian distribution, i.e., $\sigma^{2}_{\epsilon_{0}}:\mathcal{N}(0,\sigma^{2}_{\epsilon_{0}})$ and $\sigma^{2}_{\epsilon_{1}}:\mathcal{N}(0,\sigma^{2}_{\epsilon_{1}})$, where $\sigma_{\epsilon_{0}}$ ($\sigma_{\epsilon_{1}}$) is the standard deviation of the error. One can see that our OP-NHQC scheme can outperform the conventional NHQC scheme.}\label{Fig_AppC_Gaussian}
	\end{center}
\end{figure}

In actual experimental environment, the control errors in $\Omega_{0}(t)$ and $\Omega_{1}(t)$ may be random fluctuations instead of the constant variations set in the Hamiltonian.  In this way, we consider the situation that the control errors between the magnitudes of $\Omega_0(t)$ and $\Omega_1(t)$ do not change simultaneously, i.e., $\Omega_{0}(t)\rightarrow (1+\epsilon_{0}(t))\Omega_{0}(t)$ and $\Omega_{1}(t)\rightarrow (1+\epsilon_{1}(t))\Omega_{1}(t)$, where both $\epsilon_{0}(t)$ and $\epsilon_{0}(t)$ fluctuate along with time. The Hamiltonian in Eq.~\eqref{Eq:HI} with these errors can then be expressed as
\begin{equation}
	\begin{aligned}
		\mathcal{H}^{\prime}(t) =& \left[ (1+\epsilon_{0}(t))\Omega_0(t)e^{-i\phi_0(t)}|0 \rangle + (1+\epsilon_{1}(t))\Omega_1(t)e^{-i\phi_1(t)}|1\rangle \right]\langle e| + \textrm{H.c.}
	\end{aligned}
	\label{Eq:HIAP3}
\end{equation}
As shown in the main text, since the two driving fields can also be parameterized as $\Omega_0(t)=\Omega(t)\sin\frac{\theta}{2}(t)$ and $\Omega_1(t)=\Omega(t)\cos\frac{\theta}{2}(t)$, the Hamiltonian  can be rewritten as
\begin{equation}
	\begin{aligned}
		\mathcal{H}''(t) =& \left[ (1+\epsilon_{0}(t))\Omega(t)\sin\frac{\theta}{2}e^{-i\phi_0(t)}|0 \rangle \right. \\&\left.+ (1+\epsilon_{1}(t))\Omega(t)\cos\frac{\theta}{2}e^{-i\phi_1(t)}|1\rangle \right]\langle e| + \textrm{H.c.}\\
		=&(1+\epsilon(t))\Omega(t)e^{-i\phi_{0}(t)}|b_{\theta^{\prime},\phi}\rangle\langle e|+\text{H.c.}
	\end{aligned}
	\label{Eq:HIAP4}
\end{equation}
It is  clear that $\Omega(t)$ turns into $(1+\epsilon(t))\Omega(t)$ owing to the two errors. On the other hand, the error-affected bright state $|b_{\theta^{\prime},\phi}\rangle$ is
\begin{equation}
	|b_{\theta^{\prime},\phi}\rangle=\sin\frac{\theta^{\prime}}{2}|0\rangle-\cos\frac{\theta^{\prime}}{2}e^{i\phi}|1\rangle.
\end{equation}
Here, $\epsilon(t)$ and $\theta^{\prime}$ have the following forms as
\begin{equation}
	\begin{aligned}
		\epsilon(t)&=\sqrt{(1+\epsilon_{0}(t))^{2}\sin^{2}\frac{\theta}{2}+(1+\epsilon_{1}(t))^{2}\cos^{2}\frac{\theta}{2}}-1,\\
		\theta^{\prime}&=2\arctan[\frac{1+\epsilon_{0}(t)}{1+\epsilon_{1}(t)}\tan\frac{\theta}{2}].
	\end{aligned}
\end{equation}
Since we assume that $|\epsilon_{0}(t) |$, $|\epsilon_{1}(t) |\ll 1$, this implies that $\theta^{\prime}\approx \theta$,  $|b_{\theta^{\prime},\phi}\rangle \approx |b\rangle$ and $|\epsilon(t) |\ll 1$. Under these assumptions, the Hamiltonian with errors can be
\begin{equation}
	\mathcal{H}^{\epsilon}(t)=(1+\epsilon(t))\Omega(t)e^{-i\phi_{0}(t)}|b\rangle\langle e|+\text{H.c.}
\end{equation}
To test the robustness of our OP-NHQC scheme, we first simulate the fidelity of $X/2$ gate as a function of control errors $\epsilon_{0}(t)$ and $\epsilon_{1}(t)$, as shown in Fig.~\ref{Fig_A1}(b). Here, we have assumed both $\epsilon_{0}$ and $\epsilon_{1}$ are constant, which simplifies the discussion. But, it still captures the intrinsic fact that
although $\epsilon_{0}$ and $\epsilon_{1}$ do not change simultaneously, our OP-NHQC scheme demonstrates superior error resistance compared to the conventional NHQC scheme, as seen in Fig.~\ref{Fig_A1}(a).

Then, we consider the fluctuation property of the two errors. To do this, we assume the static noise model with Gaussian distribution, i.e., $\sigma^{2}_{\epsilon_{0}}:\mathcal{N}(0,\sigma^{2}_{\epsilon_{0}})$ and $\sigma^{2}_{\epsilon_{1}}:\mathcal{N}(0,\sigma^{2}_{\epsilon_{1}})$, where $\sigma_{\epsilon_{0}}$ ($\sigma_{\epsilon_{1}}$) is the standard deviation of the error \cite{Huang_fidelity_2019,ZhangChengxian_High_2020,GuoLiuJun_Optimizing_2023}.  Then,  we can average the gate fidelity over the errors, which are randomly drawn from the Gaussian distribution.  For each $\sigma_{\epsilon_{0}}$ ($\sigma_{\epsilon_{1}}$), we have considered 200 times samples. In our simulation, we have assumed $\sigma_{\epsilon_{0}}=\sigma_{\epsilon_{1}}=\sigma_{\epsilon}$. As shown in Fig.~\ref{Fig_AppC_Gaussian}, one can find that our OP-NHQC scheme still perform better than the conventional NHQC scheme, considering all the variance region.  In addition, the improvement offered by the OP-NHQC scheme becomes more and more substantial as the variance increases.








\end{document}